\documentclass[aip,apl, showpacs, reprint, floatfix]{revtex4-1}

\usepackage{amsmath}
\usepackage{graphicx}
\usepackage{textcomp}
\usepackage{url}

\begin{document}

\title{Simple remedy for spurious states in discrete ${\bf k}\cdot{\bf p}$ models of semiconductor structures}
\thanks{The authors wish to thank David Z.-Y. for directing our attention to this problem and for very helpful discussions.}

\date{\today}

\author{William R. Frensley}
\affiliation{Electrical Engineering, University of Texas at Dallas, Richardson, TX 75080}
\email{frensley@utdallas.edu}
\author{Raja N. Mir}
\affiliation{Electrical Engineering, University of Texas at Dallas, Richardson, TX 75080}

\begin{abstract}
  The spurious states found in numerical implementations of envelope function models for semiconductor heterostructures and nanostructures are
artifacts of the use of the centered-difference formula.  They are readily removed by employing a first-order
difference scheme.  The technique produces none of the loss of accuracy that is commonly feared from lower-order formulations; the fidelity
to the continuum band structure is actually improved.  Moreover the stability is absolute, thus no adjustments to the model parameters 
are required.  These properties are demonstrated in an explicit calculation of a 3-band model of a semiconductor superlattice.
\end{abstract}

\pacs{71.15.-m, 71.20.Nr, 73.21.-b}

\maketitle

\section{Introduction}

  The envelope-function formulation of Luttinger and Kohn \cite{Luttinger1955} is one of three main approaches to the 
evaluation of electron states in semiconductor heterostructures at the mesoscopic level, 
the other two being the Wannier-Slater \cite{Slater1949}
effective mass theory and empirical tight-binding models \cite{Vogl1983,Klimeck2000}.  The envelope-function approach is more
amenable to the incorporation of multi-band effects than Wannier-Slater, and its discrete implementation does not appear to 
be fixed by the underlying crystal structure in the way that tight-binding models are.  
The mesoscopic Hamiltonian in real space is thought to be straightforwardly obtained from the chosen ${\bf k}\cdot{\bf p}$
approximation in ${\bf k}$-space.  But, this procedure has been shown to produce spurious wavefunction solutions, usually 
characterized by oscillations near the maximum $k$ representable on the discrete mesh \cite{Cartoixa2003, Vogl2011}. 

  It appears that the centered-difference form of the gradient \cite{Chuang1997}  has been
universally employed.  This is presumably due to the general belief that it is both more ``accurate'' and is required to
preserve Hermiticity and physical symmetry.  We will demonstrate that the accuracy argument is incorrect, and that
the alternatives also satisfy the Hermiticity and symmetry requirements.

\section{Properties of Low-Order Difference Operators}

All continuum ${\bf k}\cdot{\bf p}$ models produce unphysical $E({\bf k})$ relations in the sense that their
asymptotic behavior is hyperbolic (or at least following some power law), rather than periodic.  
Spatially discrete models necessarily produce periodic bands, and this is a motivation to view such models as 
legitimate physical theories of mesoscopic systems in their own right, rather than mere approximations to the 
(obviously flawed) ``truth'' embodied in the continuum model. One of course wishes to exploit the available
continuum models, but only in the small-$k$ limit.  We adopt this point of view in the development
which follows.  We consider models defined on a one-dimensional discrete space with points uniformly
separated by a distance $\Delta$, so that the position of a point $z_n = n\Delta$.  

We define three different representations of the gradient $\hat{K} = -i\partial_z$:
\begin{subequations}
\begin{align}
 \left(\hat{K}_C f \right)_n &= \frac{-i}{2\Delta} \left( f_{n+1} - f_{n-1} \right) , \\
 \left(\hat{K}_L f \right)_n &= \frac{-i}{\Delta} \left( f_{n} - f_{n-1} \right) , \\
 \left(\hat{K}_R f \right)_n &= \frac{-i}{\Delta} \left( f_{n+1} - f_{n} \right) .
\end{align}
\end{subequations}
We also define the discrete Laplacian, $-\partial^2_z$ by:
\begin{equation}
 \left(\hat{L} f \right)_n = \frac{1}{\Delta^2} \left( -f_{n-1} + 2 f_n - f_{n+1} \right).
\end{equation}

These operators obey some simple relations.  The first-order differences, $\hat{K}_L$ and $\hat{K}_R$
form an adjoint pair:
\begin{equation}
 \hat{K}_L^\dagger = \hat{K}_R.
\end{equation}
Also,
\begin{equation}
  \hat{K}_L \hat{K}_R = \hat{K}_R \hat{K}_L = \hat{L},
\end{equation}
exactly for an unbounded system.  This may be readily verified by direct calculation.  On the other hand,
\begin{equation}
 \hat{K}_C^2 \neq \hat{L}.
\end{equation}
Such a discrepancy would never be tolerated in a continuum formulation (in curvilinear coordinates, for example); it 
should be an equally serious cause for concern in a discrete formulation.

\begin{figure}[tb]
 \centering
 \includegraphics[width=2.0in]{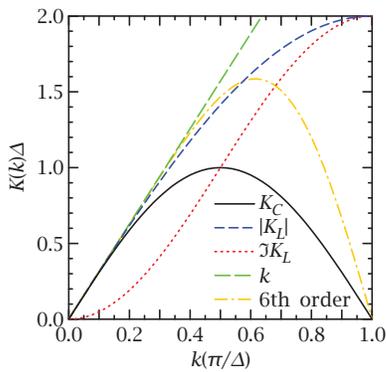}
 \caption{Dispersion relations for the different $\hat{K}$ operators.}
 \label{fig:DispRels}
\end{figure}

If we apply these operators to a plane-wave state $f_n = e^{ink\Delta}$ we obtain the respective dispersion relations
$K(k)$:
\begin{subequations} \label{eqn:KdispRels}
\begin{align}
 {K}_C(k) &= \frac{1}{\Delta} \sin(k\Delta) , \\
 {K}_L(k) &= \frac{-i}{\Delta} \left( 1 - e^{-ik\Delta} \right) , \\
 {K}_R(k) &= \frac{i}{\Delta} \left( 1 - e^{ik\Delta} \right) .
\end{align}
\end{subequations}
Each of these functions has the same real part, equal to $K_C(k)$, but $K_L$ and $K_R$ also have imaginary parts.  
The magnitude of those $K$s is particularly significant:
\begin{equation}
 |K_L| = |K_R| = \sqrt{K_L K_R} = \frac{1}{\Delta} \sqrt{2 - 2\cos(k\Delta)} .
\end{equation}
These functions are plotted in Fig.\ \ref{fig:DispRels}.
Observe that $|K_{L,R}|$ is closer to the straight line representing $k$ than is $K_C$, implying that the first-order scheme
is actually {\it more} accurate than the second-order one.

\section{Application to the Simple Two-Band Model}

\begin{figure}[tb]
 \centering
 \includegraphics[width=2.3in]{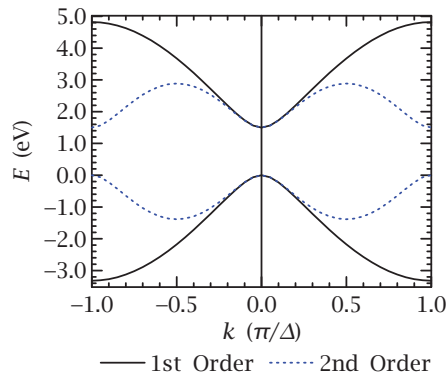}
 \caption{Discrete band structures for the simple 2-band ${\bf k}\cdot{\bf p}$ model using first and second order discretizations.  
 Note how the band structures directly reflect the dispersion relations of Fig.\ \ref{fig:DispRels}.}
 \label{fig:kp2band}
\end{figure}

We can quickly see how the dispersion relations (\ref{eqn:KdispRels}) influence the band structure of the discrete
model by applying the respective operators to the simple two-band ${\bf k}\cdot{\bf p}$ model.  The continuum Hamiltonian
\begin{gather}
 \hat{H} = \begin{bmatrix} E_C & Pk \\ Pk & E_V \end{bmatrix} 
\intertext{ becomes }
  \Hat{H}_2 =  \begin{bmatrix} E_C & \hat{K}_C P \\ P\hat{K}_C & E_V \end{bmatrix}
\intertext{ in the usual second order formulation, but the first order model gives }
  \hat{H}_1 =  \begin{bmatrix}  E_C & \hat{K}_R P \\ P\hat{K}_L & E_V \end{bmatrix} .  \label{eqn:Hleft}
\end{gather}
Applying any of these forms to a plane wave yields the general bandstructure:
\begin{equation} \label{eqn:gen2bandEk}
 E(k) = \frac{E_C+E_V}{2} \pm \sqrt{ \left( \frac{E_C+E_V}{2} \right)^2 + \left| K \right|^2 }.
\end{equation}
Inserting the appropriate dispersion relations into (\ref{eqn:gen2bandEk}) produces the results shown in
Fig.\ \ref{fig:kp2band}.

The second-order model shows the well-known nonmonotonic behavior which places spurious high-$k$ solutions within the energy ranges
of interest in heterostructure applications.  The energy bands produced by the first-order model are monotonic; thus there will be
no spurious solutions.  The band structures clearly reflect the underlying difference-operator dispersion relations.

\section{Realistic Three-Band Model}

\begin{figure}[tb]
 \centering
 \includegraphics[width=2.5in]{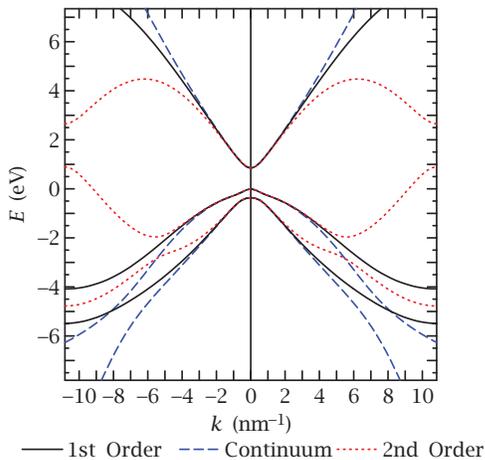}
 \caption{Band structure resulting from the first-order and centered-diference models, compared to the continuum solution.  The 
 Hamiltonian is the 3-band model employed by Cartoix\`a applied to In$_{0.53}$Ga$_{0.47}$As with a mesh spacing of 0.29 nm which equals $a/2$.}
 \label{fig:InGaAs3band}
\end{figure}

To demonstrate the effect of the first-order difference formulation, we have applied it to the three-band ${\bf k}\cdot{\bf p}$ model
in which Cartoix\`a, Ting and McGill (CTM) identified the origin of spurious solutions \cite{Cartoixa2003}.  In the Hamiltonian (eq.\ 6 of 
that paper) we replace factors of $k_z$ with $\hat{K}_L$ if they appear below the diagonal, or with $\hat{K}_R$ if they
appear above the diagonal, to maintain the Hermiticity of the Hamiltonian. 
 Factors of the form $\gamma k^2_z $ are replaced by the usual three-point formula for the Laplacian
 wherever they appear.  The resulting dispersion relations are shown in Fig.\ \ref{fig:InGaAs3band}, along with
the continuum results and those of the usual centered-difference discretization.  The nonmonotonic bands leading to spurious solutions
are present in the centered-difference results, as reported in CTM, but are eliminated by the first-order difference.
Again, the shapes of the dispersion relations of Fig.\ \ref{fig:DispRels} appear clearly in the respective band structures.

\section{Heterostructures}

  When we apply the discrete Hamiltonian to its intended purpose, solution of spatially-resolved heterostructures, we obtain a 
block-tridiagonal matrix (neglecting the corner blocks that implement periodic boundary conditions).  The one-sided differences 
lead to an asymmetry in the structure of this matrix in the diagonal direction, but we will show that this need not break the 
physical symmetry of the underlying system. There are two possible conventions: to place $\hat{K}_L$ operators in the lower
triangle, or to place $\hat{K}_R$ operators there.  We will for the moment adopt the first alternative.

When the system includes heterojunctions, the interband parameters $P$ and the mass
parameters $\gamma$ become position-dependent, represented by diagonal matrices, and we have to take care with the operator ordering
as indicated in (\ref{eqn:Hleft}).  This places elements of value $P_n$ in off-diagonal elements of the $\hat{H}_{n, n}$
diagonal block, and in the $\hat{H}_{n, n-1}$ and $\hat{H}_{n-1, n}$ off-diagonal blocks.  The value $P_{n-1}$ will appear
in the $\hat{H}_{n-1, n-1}$ diagonal block.  This means that the material at point $n$ must be assumed to extend completely 
across the $(n-1, n)$ interval, and if the material at $n-1$ is different, the heterojunction must be coincident with the
point at $n-1$.  This is in contrast to the usual assumption that the junction lies at the center of the interval between 
points. \cite{Frensley1994} 

This mapping between the physical structure and the discrete model uniquely determines the rest of the design of the 
discretization.  The band energies on the diagonal of $\hat{H}_{n-1, n-1}$ of must be set to the average of the two
materials: \textonehalf$(E_{n-1} + E_n)$.  The terms containing $k^2$ are defined as:
\begin{equation*}
 \gamma k^2 \rightarrow \hat{K}_L \hat{\gamma} \hat{K}_R,  
\end{equation*}
leading to:
\begin{multline}
 \Delta^2 (\hat{K}_L \hat{\gamma} \hat{K}_R f)_{n-1} \\ 
 		= -\gamma_{n-1} f_{n-2} + (\gamma_{n-1} + \gamma_n) f_{n-1} - \gamma_n f_{n} .
\end{multline}

If we adopt the other convention, interchanging the $\hat{K}_L$ and $\hat{K}_R$ operators, we
can make a similar argument that places the heterojunction coincident with the $n+1$ meshpoint,
if the material changes between points $n$ and $n+1$.  The net effect is that the mapping betweeen
the physical structure and the discrete model must be translated by one mesh interval.  We thus
find a rather novel type of symmetry: the formulation is invariant under the interchange of left
and right-hand differences, provided that we also perform the necessary translation of the discrete
model.  

To demonstrate the effectiveness of present formulation, we have applied it to a system similar to that 
studied by CTM, using the same 3-band model.  The system consists of 15 intervals of InP, 20 intervals of In$_{0.53}$Ga$_{0.47}$As, and 15 intervals of InP,
with $\Delta = 0.29$ nm, as before.  Periodic boundary conditions are applied, so that the resulting eigenenergies and
states represent those at the miniband edges of a superlattice.  There are a total of 50 mesh points yielding a matrix of
dimension 150 for the 3-band model. 
Solutions were calculated for the left-hand, right-hand
and centered-difference models of this system.  The left- and right-hand models show no evidence of spurious states. 
The energies from the two one-sided models agree to 10 decimal places, and show no states lying in the bandgap of InGaAs. 

\begin{figure}[tb]
 \centering
 \includegraphics[width=2.5in]{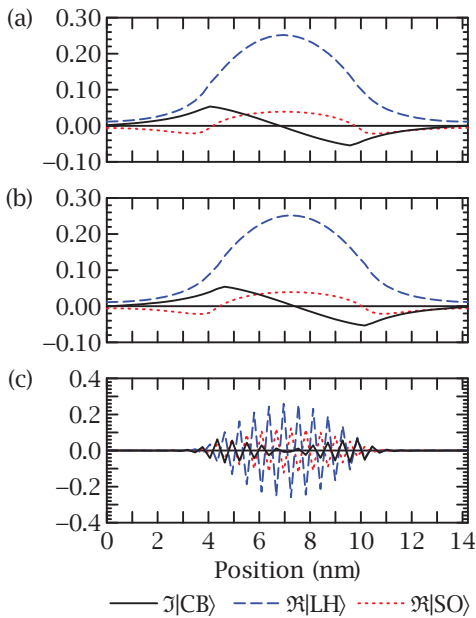}
 \caption{The 100th state from the superlattice calculation described in the text, which should be the highest-energy light-hole state.  
 The discrete formulations are:
 (a) left-hand, (b) right-hand and (c) centered difference. Case (c) is clearly spurious.  }
 \label{fig:SL3band}
\end{figure} 
The wavefunctions produced by the three models for the 100th eigenstate are shown in Fig.\ \ref{fig:SL3band}.  This should
be the highest-lying light hole state confined in the quantum well.  The one-sided models place this state 0.076 eV below
the InGaAs valence band edge, while the centered-difference model places it 0.79 eV above that edge, just below the 
conduction band, and its wavefunction clearly indicates that it is a result of the unphysical light-hole dispersion 
relation of this model.  A comparison of the wavefunctions from the left- and right-hand models indicates that their
symmetry is not as exact as that of the energies, but they represent functions that are displaced by something between
one and two mesh intervals.  

An examination of the other states in the general vicinity of the bandgap reveals that their properties are fully consistent
with the expectations derived from the bulk dispersion relations of Fig.\ \ref{fig:InGaAs3band}.  The wavefunctions
from the left- and right-hand models show no evidence whatsoever of high-$k$ oscillation.  All the centered-difference
solutions show such effects, except those within the energy range of 0.9--2.5 eV where there is a gap in the high-$k$
region.  The dispersion relations do indeed determine the outcome.

In view of these results, we examined the dispersion
relations for a number of common cubic III-V compounds. We used the parameters tabulated by Vurgraftman, Meyer, and Ram-Mohan\cite{Vurgaftman2001} as applied
to the 3-band model of CTM.  We find that the first-order formulations always produce
monotonic band dispersions and will thus be free from spurious solutions.  The centered-difference formulation aways shows 
retrograde regions in some energy range.

The first-order scheme thus appears to be entirely robust, if the continuum formulation does not produce spurious states.  
Some continuum ${\bf k}\cdot{\bf p}$ formulations using larger basis sets do admit spurious solutions.\cite{Yang2005}  
The effect of the first-order dispersion relations on such models has not yet been investigated.

\section{Discussion}

  Attempts to remove the spurious solutions have focused on using ambiguities in the ${\bf k}\cdot{\bf p}$ parameters to move the unphysical band minima to 
more remote energies \cite{Cartoixa2003}, or to construct more elaborate formalisms to better approximate the continuum behavior \cite{Vogl2011}.
While the centered-difference has been explicitly identified with the problem, the success that asymmetric differences have shown in other 
contexts has been overlooked.  In fluid dynamics the term variously known as drift, advection, or streaming can be treated with absolute stability
using a first-order upwind difference scheme \cite{Roache1998}. If centered differences or a hybrid scheme is used, the stability is very 
much dependent upon the parameters of the problem.   In the development of time-dependent quantum transport simulations in semiconductors,\cite{FrensleyRMP1990}
it was found that a Wigner-function representation (which yields a drift term) combined with upwind differencing was the stable formulation.

 The reason for the failure of antisymmetric (as opposed to asymmetric) formulations can be fairly easily explained, but we can only sketch the outline of the
 argument here.  Antisymmetric difference formulas produce purely real dispersion relations, which are necessarily periodic.  Thus, in addition to the
 true zero at the origin there must also be a zero at $k = \pm\pi/\Delta$, and the dispersion relation in the vicinity of that zero is the source of
 the spurious solutions.  One can never remove this effect from an antisymmetric formulation, as illustrated by the 6th-order dispersion curve 
 shown in Fig.\ \ref{fig:DispRels}.  An asymmetric difference gives a complex-valued dispersion, which can avoid the spurious zero by arcing through
 the complex plane, as we see for the first-order differences.  As the mesh spacing is reduced the antisymmetric operators will converge non-uniformly
due to the persistence of the spurious components.  The asymmetric formulations will converge uniformly
 to the continuum operator.
 
   The above discussion is offered as an approach to understanding the inevitablility of the present results, which directly
contradict the conventional understanding.  We are {\it not} asserting that the validity of this formulation depends upon its  
behavior as the continuum limit is approached.  The validity of a discrete model depends upon its own properties and on the
degree to which its results conform to the known behavior of the system being simulated. 

In conclusion, let us quote Eissfeller and Vogl \cite{Vogl2011}: ``(The centered difference) scheme has been identified earlier to be the cause of unphysical
solutions for many semiconductor nanostructures, but no easy remedy is known.''  Their objection to first-order schemes is that these would break the symmetry
of the Hamiltonian.  We have shown that a distinction must be made between mathematical symmetry and physical symmetry, and the latter can be preserved with 
proper attention to the details of the discretization.  Thus, the first-order difference is precisely that easy remedy 
which has been sought.

\bibliography{FrensleyEnvFn2015x2}

\begin{thebibliography}{12}%
\makeatletter
\providecommand \@ifxundefined [1]{%
 \@ifx{#1\undefined}
}%
\providecommand \@ifnum [1]{%
 \ifnum #1\expandafter \@firstoftwo
 \else \expandafter \@secondoftwo
 \fi
}%
\providecommand \@ifx [1]{%
 \ifx #1\expandafter \@firstoftwo
 \else \expandafter \@secondoftwo
 \fi
}%
\providecommand \natexlab [1]{#1}%
\providecommand \enquote  [1]{``#1''}%
\providecommand \bibnamefont  [1]{#1}%
\providecommand \bibfnamefont [1]{#1}%
\providecommand \citenamefont [1]{#1}%
\providecommand \href@noop [0]{\@secondoftwo}%
\providecommand \href [0]{\begingroup \@sanitize@url \@href}%
\providecommand \@href[1]{\@@startlink{#1}\@@href}%
\providecommand \@@href[1]{\endgroup#1\@@endlink}%
\providecommand \@sanitize@url [0]{\catcode `\\12\catcode `\$12\catcode
  `\&12\catcode `\#12\catcode `\^12\catcode `\_12\catcode `\%12\relax}%
\providecommand \@@startlink[1]{}%
\providecommand \@@endlink[0]{}%
\providecommand \url  [0]{\begingroup\@sanitize@url \@url }%
\providecommand \@url [1]{\endgroup\@href {#1}{\urlprefix }}%
\providecommand \urlprefix  [0]{URL }%
\providecommand \Eprint [0]{\href }%
\providecommand \doibase [0]{http://dx.doi.org/}%
\providecommand \selectlanguage [0]{\@gobble}%
\providecommand \bibinfo  [0]{\@secondoftwo}%
\providecommand \bibfield  [0]{\@secondoftwo}%
\providecommand \translation [1]{[#1]}%
\providecommand \BibitemOpen [0]{}%
\providecommand \bibitemStop [0]{}%
\providecommand \bibitemNoStop [0]{.\EOS\space}%
\providecommand \EOS [0]{\spacefactor3000\relax}%
\providecommand \BibitemShut  [1]{\csname bibitem#1\endcsname}%
\let\auto@bib@innerbib\@empty
%</preamble>
\bibitem [{\citenamefont {Luttinger}\ and\ \citenamefont
  {Kohn}(1955)}]{Luttinger1955}%
  \BibitemOpen
  \bibfield  {author} {\bibinfo {author} {\bibfnamefont {J.~M.}\ \bibnamefont
  {Luttinger}}\ and\ \bibinfo {author} {\bibfnamefont {W.}~\bibnamefont
  {Kohn}},\ }\href {\doibase 10.1103/PhysRev.97.869} {\bibfield  {journal}
  {\bibinfo  {journal} {Phys. Rev.}\ }\textbf {\bibinfo {volume} {97}},\
  \bibinfo {pages} {869} (\bibinfo {year} {1955})}\BibitemShut {NoStop}%
\bibitem [{\citenamefont {Slater}(1949)}]{Slater1949}%
  \BibitemOpen
  \bibfield  {author} {\bibinfo {author} {\bibfnamefont {J.~C.}\ \bibnamefont
  {Slater}},\ }\href {\doibase 10.1103/PhysRev.76.1592} {\bibfield  {journal}
  {\bibinfo  {journal} {Phys. Rev.}\ }\textbf {\bibinfo {volume} {76}},\
  \bibinfo {pages} {1592} (\bibinfo {year} {1949})}\BibitemShut {NoStop}%
\bibitem [{\citenamefont {Vogl}, \citenamefont {Hjalmarson},\ and\
  \citenamefont {Dow}(1983)}]{Vogl1983}%
  \BibitemOpen
  \bibfield  {author} {\bibinfo {author} {\bibfnamefont {P.}~\bibnamefont
  {Vogl}}, \bibinfo {author} {\bibfnamefont {H.~P.}\ \bibnamefont
  {Hjalmarson}}, \ and\ \bibinfo {author} {\bibfnamefont {J.~D.}\ \bibnamefont
  {Dow}},\ }\href {\doibase http://dx.doi.org/10.1016/0022-3697(83)90064-1}
  {\bibfield  {journal} {\bibinfo  {journal} {J.\ Phys.\ Chem.\ Solids}\
  }\textbf {\bibinfo {volume} {44}},\ \bibinfo {pages} {365} (\bibinfo {year}
  {1983})}\BibitemShut {NoStop}%
\bibitem [{\citenamefont {Klimeck}\ \emph {et~al.}(2000)\citenamefont
  {Klimeck}, \citenamefont {Bowen}, \citenamefont {Boykin},\ and\ \citenamefont
  {Cwik}}]{Klimeck2000}%
  \BibitemOpen
  \bibfield  {author} {\bibinfo {author} {\bibfnamefont {G.}~\bibnamefont
  {Klimeck}}, \bibinfo {author} {\bibfnamefont {R.~C.}\ \bibnamefont {Bowen}},
  \bibinfo {author} {\bibfnamefont {T.~B.}\ \bibnamefont {Boykin}}, \ and\
  \bibinfo {author} {\bibfnamefont {T.~A.}\ \bibnamefont {Cwik}},\ }\href
  {\doibase http://dx.doi.org/10.1006/spmi.2000.0862} {\bibfield  {journal}
  {\bibinfo  {journal} {Superlattices and Microstructures}\ }\textbf {\bibinfo
  {volume} {27}},\ \bibinfo {pages} {519 } (\bibinfo {year}
  {2000})}\BibitemShut {NoStop}%
\bibitem [{\citenamefont {Cartoix\`a}, \citenamefont {Ting},\ and\
  \citenamefont {McGill}(2003)}]{Cartoixa2003}%
  \BibitemOpen
  \bibfield  {author} {\bibinfo {author} {\bibfnamefont {X.}~\bibnamefont
  {Cartoix\`a}}, \bibinfo {author} {\bibfnamefont {D.~Z.-Y.}\ \bibnamefont
  {Ting}}, \ and\ \bibinfo {author} {\bibfnamefont {T.~C.}\ \bibnamefont
  {McGill}},\ }\href@noop {} {\bibfield  {journal} {\bibinfo  {journal} {J.\
  Appl.\ Phys.}\ }\textbf {\bibinfo {volume} {93}} (\bibinfo {year}
  {2003})}\BibitemShut {NoStop}%
\bibitem [{\citenamefont {Eissfeller}\ and\ \citenamefont
  {Vogl}(2011)}]{Vogl2011}%
  \BibitemOpen
  \bibfield  {author} {\bibinfo {author} {\bibfnamefont {T.}~\bibnamefont
  {Eissfeller}}\ and\ \bibinfo {author} {\bibfnamefont {P.}~\bibnamefont
  {Vogl}},\ }\href {\doibase 10.1103/PhysRevB.84.195122} {\bibfield  {journal}
  {\bibinfo  {journal} {Phys. Rev. B}\ }\textbf {\bibinfo {volume} {84}},\
  \bibinfo {pages} {195122} (\bibinfo {year} {2011})}\BibitemShut {NoStop}%
\bibitem [{\citenamefont {Chuang}\ and\ \citenamefont
  {Chang}(1997)}]{Chuang1997}%
  \BibitemOpen
  \bibfield  {author} {\bibinfo {author} {\bibfnamefont {S.~L.}\ \bibnamefont
  {Chuang}}\ and\ \bibinfo {author} {\bibfnamefont {C.~S.}\ \bibnamefont
  {Chang}},\ }\href {http://stacks.iop.org/0268-1242/12/i=3/a=004} {\bibfield
  {journal} {\bibinfo  {journal} {Semicond.\ Sci.\ Technol.}\ }\textbf
  {\bibinfo {volume} {12}},\ \bibinfo {pages} {252} (\bibinfo {year}
  {1997})}\BibitemShut {NoStop}%
\bibitem [{\citenamefont {Frensley}(1994)}]{Frensley1994}%
  \BibitemOpen
  \bibfield  {author} {\bibinfo {author} {\bibfnamefont {W.~R.}\ \bibnamefont
  {Frensley}},\ }\enquote {\bibinfo {title} {Quantum transport},}\ in\
  \href@noop {} {\emph {\bibinfo {booktitle} {Heterostructures and Quantum
  Devices}}}\ (\bibinfo  {publisher} {Academic Press},\ \bibinfo {year}
  {1994})\ pp.\ \bibinfo {pages} {273--303}\BibitemShut {NoStop}%
\bibitem [{\citenamefont {Vurgaftman}, \citenamefont {Meyer},\ and\
  \citenamefont {Ram-Mohan}(2001)}]{Vurgaftman2001}%
  \BibitemOpen
  \bibfield  {author} {\bibinfo {author} {\bibfnamefont {I.}~\bibnamefont
  {Vurgaftman}}, \bibinfo {author} {\bibfnamefont {J.~R.}\ \bibnamefont
  {Meyer}}, \ and\ \bibinfo {author} {\bibfnamefont {L.~R.}\ \bibnamefont
  {Ram-Mohan}},\ }\href {\doibase http://dx.doi.org/10.1063/1.1368156}
  {\bibfield  {journal} {\bibinfo  {journal} {Journal of Applied Physics}\
  }\textbf {\bibinfo {volume} {89}},\ \bibinfo {pages} {5815} (\bibinfo {year}
  {2001})}\BibitemShut {NoStop}%
\bibitem [{\citenamefont {Yang}\ and\ \citenamefont {Chang}(2005)}]{Yang2005}%
  \BibitemOpen
  \bibfield  {author} {\bibinfo {author} {\bibfnamefont {W.}~\bibnamefont
  {Yang}}\ and\ \bibinfo {author} {\bibfnamefont {K.}~\bibnamefont {Chang}},\
  }\href {\doibase 10.1103/PhysRevB.72.233309} {\bibfield  {journal} {\bibinfo
  {journal} {Phys. Rev. B}\ }\textbf {\bibinfo {volume} {72}},\ \bibinfo
  {pages} {233309} (\bibinfo {year} {2005})}\BibitemShut {NoStop}%
\bibitem [{\citenamefont {Roache}(1998)}]{Roache1998}%
  \BibitemOpen
  \bibfield  {author} {\bibinfo {author} {\bibfnamefont {P.~J.}\ \bibnamefont
  {Roache}},\ }\href@noop {} {\emph {\bibinfo {title} {Fundamentals of
  Computational Fluid Dynamics}}}\ (\bibinfo  {publisher} {Hermosa
  Publishers},\ \bibinfo {address} {Albuquerque, NM},\ \bibinfo {year} {1998})\
  p.~\bibinfo {pages} {5}\BibitemShut {NoStop}%
\bibitem [{\citenamefont {Frensley}(1990)}]{FrensleyRMP1990}%
  \BibitemOpen
  \bibfield  {author} {\bibinfo {author} {\bibfnamefont {W.~R.}\ \bibnamefont
  {Frensley}},\ }\href {\doibase 10.1103/RevModPhys.62.745} {\bibfield
  {journal} {\bibinfo  {journal} {Rev. Mod. Phys.}\ }\textbf {\bibinfo {volume}
  {62}},\ \bibinfo {pages} {745} (\bibinfo {year} {1990})}\BibitemShut
  {NoStop}%
\end{thebibliography}%

\end{document}